# Leaky-wave metasurfaces for integrated photonics

Heqing Huang[1,†], Adam C. Overvig[1,2,†], Yuan Xu[1], Stephanie C. Malek[1], Cheng-Chia Tsai[1], Andrea Alù[2,3,*], and Nanfang Yu[1,*]

[1]Department of Applied Physics and Applied Mathematics, Columbia University, New York, NY 10027, USA.

[2]Photonics Initiative, Advanced Science Research Center, City University of New York, New York, NY 10031, USA

[3]Physics Program, Graduate Center of the City University of New York, New York, NY 10016, USA

†Authors contributed equally

*Corresponding authors: ny2214@columbia.edu, aalu@gc.cuny.edu

## Abstract

*Metasurfaces have been rapidly advancing our command over the many degrees of freedom of light within compact, lightweight devices. However, so far, they have mostly been limited to manipulating light in free space. Grating couplers provide the opportunity of bridging far-field optical radiation and in-plane guided waves, and thus have become fundamental building blocks in photonic integrated circuits. However, their operation and degree of light control is much more limited than metasurfaces. Metasurfaces integrated on top of guided wave photonic systems have been explored to control the scattering of light off-chip with enhanced functionalities – namely, point-by-point manipulation of amplitude, phase or polarization. However, these efforts have so far been limited to controlling one or two optical degrees of freedom at best, and to device configurations much more complex compared to conventional grating couplers. Here, we introduce leaky-wave metasurfaces, which are based on symmetry-broken photonic crystal slabs that support quasi-bound states in the continuum. This platform has a compact form factor equivalent to the one of conventional grating couplers, but it provides full command over amplitude, phase and polarization (four optical degrees of freedom) across large apertures. We present experimental demonstrations of various*

*functionalities for operation at λ = 1.55 μm based on leaky-wave metasurfaces, including devices for phase and amplitude control at a fixed polarization state, and devices controlling all four optical degrees of freedom. Our results merge the fields of guided and free-space optics under the umbrella of metasurfaces, exploiting the hybrid nature of quasi-bound states in the continuum, for opportunities to advance in disruptive ways imaging, communications, augmented reality, quantum optics, LIDAR and integrated photonic systems.*

## Introduction

A monochromatic optical wavefront in free-space is characterized by four degrees of freedom at each point in space, $(A, \Phi, \psi, \chi)$: its amplitude $A$, phase $\Phi$, and polarization state, with elliptical parameters $\psi$ and $\chi$ representing polarization orientation and ellipticity, respectively. Manipulation of these degrees of freedom is among the key goals of contemporary photonics research. Metasurfaces [1]-[3] – flat optical devices composed of arrays of subwavelength scatterers – have been offering a flexible and powerful platform for producing desired wavefronts starting from unpatterned plane waves incident from free space, effectively compactifying table-top optical setups into multifunctional thin films [4]. Metasurfaces at optical frequencies have been widely used to spatially manipulate phase, but have also been shown to manipulate amplitude and phase [5],[6], phase and polarization state [7],[8], and recently all four parameters simultaneously [9]-[12] and beyond [13]. These concepts can be leveraged to an even larger extent in the radio-frequency (RF) spectrum, for which multi-layered fabrication and the large conductivity of metals enable exquisite and deeply subwavelength control of electromagnetic radiation [14]-[19]. In addition to free-space excitation, RF leaky-wave antennas [20]-[22] have been developed over several decades [23] to produce free-space beams by scattering radiation originating from guided modes. Metasurface concepts have recently advanced this field [19],[24]-[27], but these approaches are not straightforwardly transferable to optical frequencies. For comparison, grating couplers (GCs) in integrated photonics also generate free-space light from in-plane guided sources, but are largely limited in controlling the optical degrees of freedom ($A, \Phi, \psi, \chi$) and their spatial profile.

Recent years have seen a rapidly growing interest in incorporating metasurface principles into integrated photonics [28],[29] and, very recently, in generating wavefronts from in-plane guided modes [30]-[35]. This capability is of great interest to the broader optics community, representing a novel opportunity to compactify off-chip emission of customized free-space wavefronts, while also leveraging on-chip command of light based on the commercially maturing field of photonic integrated circuits (PICs). The customizability of a metasurface-based replacement for GCs offers exciting opportunities for optical communications, augmented reality, quantum optics, and LIDAR. However, so far, the presented approaches offer only partial solutions, not capable of fully commanding the coupling of guided waves to far-field radiation. At most, two optical degrees of freedom have been manipulated at once, limiting applications to scalar waves (see **Supplementary Table S.1** for recent progress in this context).

Additionally, contrary to corrugated structures typically seen in GCs used in integrated photonics [**Fig. 1(a,b)**], the structures proposed so far are composed of a waveguiding layer and a metasurface as two separate objects [**Fig. 1(c,d)**], which hinders integrability, scalability and compactness. In early examples, separated metasurface and GC layers were used [36], while in more recent examples the metasurface was placed in the evanescent field of the guided mode to both scatter and manipulate the phase profile. Both metallic [31] and dielectric [33] structures have been explored, introducing either optical loss or high-aspect-ratio dielectric structures typical of metasurface approaches. Adding such lossy or high-aspect-ratio metasurface layer on top of existing waveguiding structures complicates its implementation in comparison to conventional GCs. Additionally, so far these efforts have been limited to small surface emission apertures. These factors hinder the adoption of this approach in PICs. In contrast, a device configuration featuring the compact form factor typical of GCs, and capable of robust, subwavelength control of all four degrees of freedom of light $(A, \Phi, \psi, \chi)$ introduces a universal generalization (and where appropriate, replacement) of GCs, advancing existing approaches in both form and function.

In this work, we introduce a leaky-wave metasurface (LWM) platform, based on weakly corrugated, symmetry-broken photonic crystal slabs, that supports a quasi-bound wave capable of arbitrarily tailoring the scattered field $(A, \Phi, \psi, \chi)$ with subwavelength resolution. LWMs inherit the form of GCs, while greatly improving on the functionality of metasurface-on-waveguide solutions [**Fig. 1(e,f)**]. We experimentally implement the proposed concepts in the near-infrared (near $\lambda=1.55\mu m$) based on a silicon nitride and polymer system, wherein nanostructured polymer zones on an unpatterned silicon nitride thin film define both rib waveguides and LWMs. The design principles are rooted in quasi-bound states in the continuum [37],[38] and diffractive nonlocal metasurfaces [38]-[42], enabling a rational design approach with largely independent mapping of four geometric parameters to the four optical degrees of freedom. Full-wave simulations are used to create a library of meta-units, each one composed of two staggered rows of ellipse dimers. With simple corrections based on the propagation of the guided mode, reference to this library specifies in a rational way the full geometry of the LWM based on desired spatial profiles of amplitude, phase and polarization of free-space emission. To demonstrate the flexibility of our platform, we realize focused emission of a desired linear polarization (with wavelength-tuned scanning of the focal spot), a vortex beam generated in concert with a Gaussian reference beam, a two-image hologram encoded in the amplitude and phase of a single polarization, a four-image hologram encoded in the amplitudes and phases of two orthogonal polarizations, and a converging Poincaré beam [43].

Because our platform is designed based on symmetry breaking principles, it exemplifies a universal approach for controlling the leakage of guided waves for a wide range of material systems (e.g., metals, dielectrics, 2D materials) and wave phenomena (e.g., RF, acoustics, elastics, surface waves). Notably, with simple adjustments our approach is compatible with conventional integrated photonic architectures, such as etched waveguides and silicon-on-insulator wafers. The methods demonstrated herein can therefore be readily applied to integrated photonic systems, and open a variety of avenues for future research to bridge guided and open systems, ubiquitous across several scientific disciplines.

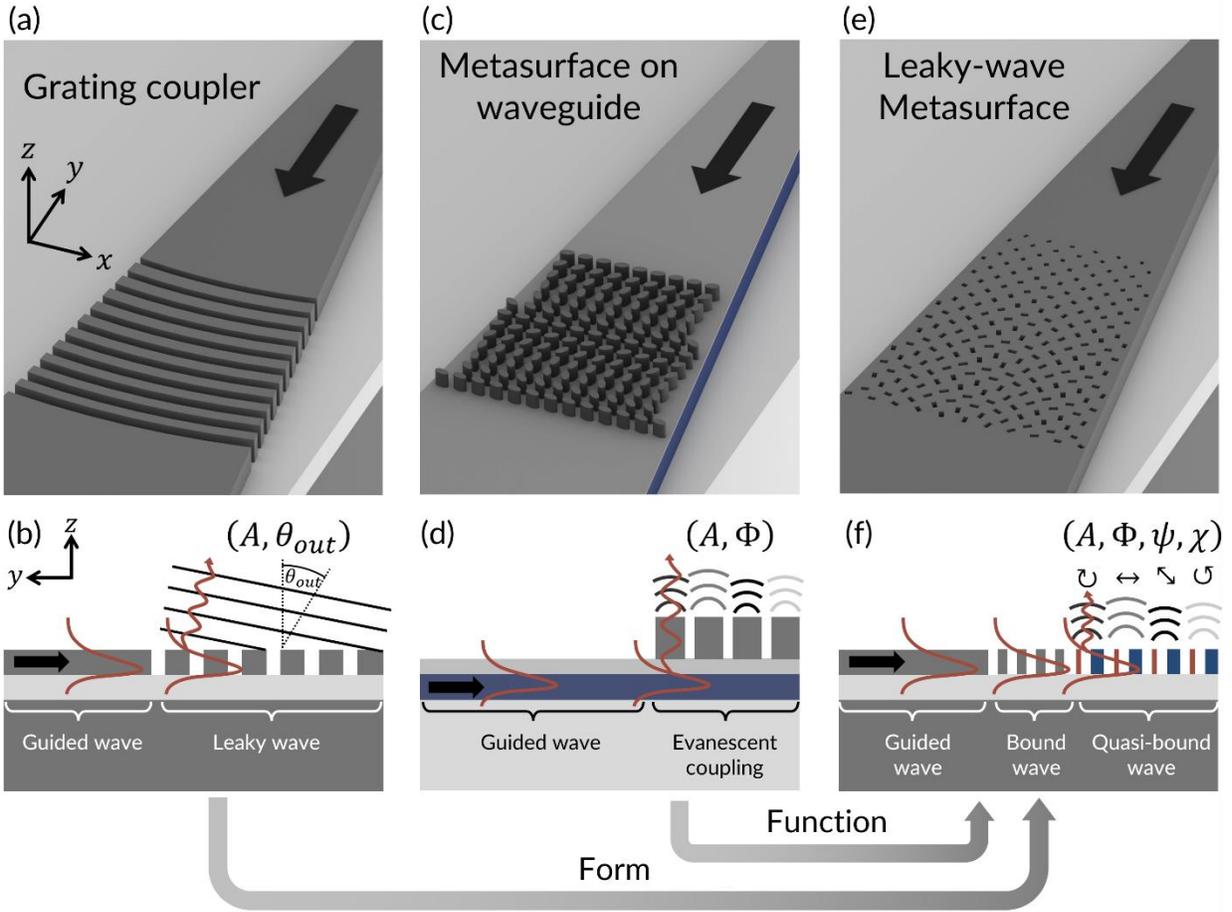

**Figure 1**. Out-coupling in integrated photonic devices. (a) Schematic depiction of a typical GC based on a fully-etched waveguide grating. (b) Side view showing the functionality of a conventional GC, wherein the duty cycle and period may be changed to alter the amplitude and outgoing angle. (c) Schematic depiction of a metasurface-on-waveguide (MOW) approach, here based on a high-index dielectric pillar array on a waveguide layer (blue). (d) Side view showing the functionality of a MOW, where the metasurface scatters the evanescent component of a guided wave into the far-field, typically limited to controlling two optical parameters of the surface emission. (e) Schematic depiction of the leaky-wave metasurface (LWM) platform introduced in this work, in which a perturbed (i.e., symmetry-broken) subwavelength photonic crystal slab supports a tailored quasi-bound wave controlling off-chip coupling. (f) Side view showing that the LWM has a compact form equivalent to a GC, but more advanced functionality compared to a MOW, offering simultaneous control over all four degrees of freedom of light $(A, \Phi, \psi, \chi)$.

## Operating principles and metasurface design

The key operating principle of our LWM platform is the deliberate perturbation of a guided mode supported by a periodic structure with subwavelength pitch (i.e., a bound wave under the light line) into a quasi-bound wave (above the light line). As sketched in **Fig. 1(f)**, a guided mode incident from a waveguide couples to a bound wave in the subwavelength periodic structure, and then leaks to free space due to a period-doubling perturbation [44]. For a proof-of-principle, we use the configuration shown in **Figs. 2(a,b)** based on a rib waveguide and a metasurface defined within a thin layer of polymer ($n\sim1.48$) atop an unpatterned thin film of silicon nitride ($n\sim2.0$) sitting on a silicon dioxide substrate ($n\sim1.44$) (see Methods for detailed geometrical parameters). The metasurface in its unperturbed state [**Fig. 2(c)**] is a two-dimensional photonic crystal composed of a rectangular lattice of circular holes with pitches $a_x$ and $a_y$; it supports a bound wave traveling in the $-y$ direction, whose effective wavelength is approximately $\lambda_{eff} \approx 2a_y$. Two independent perturbations are applied to the top pair of circular holes [**Fig. 2(d)**, Perturbation 1] and to the bottom pair of circular holes [**Fig. 2(e)**, Perturbation 2]. These perturbations double the effective lattice pitches to $2a_x$ and $2a_y$, modifying the first Brillouin zone (FBZ) of the unperturbed lattice [**Fig. 2(f)**] and its band diagram [**Fig. 2(g)**] into the zone-folded versions shown in **Figs. 2(h,i)**. The resulting band structure supports transverse-magnetic (TM) modes near $\lambda_{eff} \approx 2a_y$ in the form of a Dirac point at normal incidence, and red and black arrows in **Figs. 2(g,i)** track example states before and after the perturbation. These arrows span the Γ point of the perturbed band structure, enabling operation anywhere at or near normal to the device plane [45]. We note that an undesirable flat band also arises, degenerate with the Dirac point when $a_x = a_y$, which may be blueshifted or redshifted by detuning from this condition, if desired (see **Supplementary Materials S.2**). In this way, the scheme shown in **Fig. 2** yields a subwavelength lattice that scatters light to free space at or near normal to the device plane, a process exclusively controlled by a geometric perturbation: deliberate engineering of this symmetry-breaking perturbation determines both *if* and *how* the wave leaks to free space, pixel by pixel across the LWM aperture.

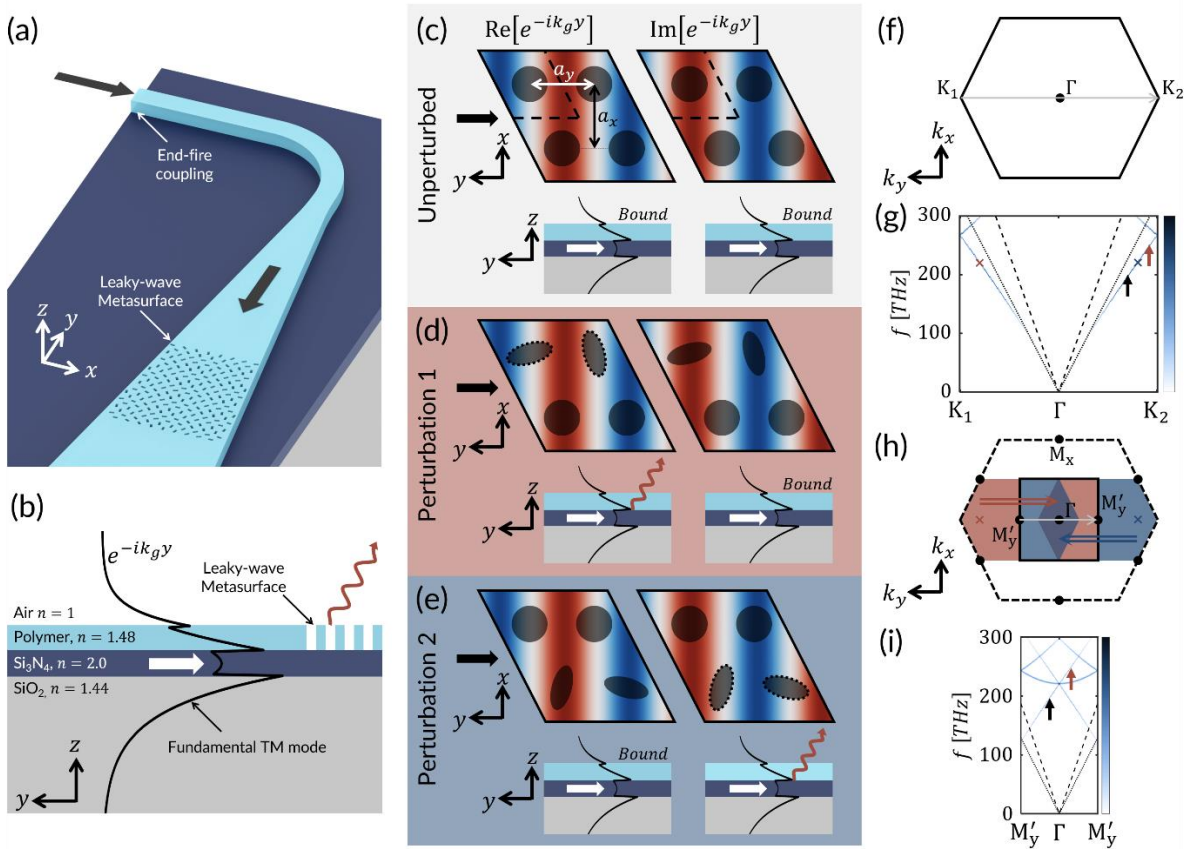

**Figure 2.** Perturbative scheme for rationally designed LWMs. (a) Schematic and (b) side view of the device geometry. (c-e) Perturbative scheme for simultaneous control of the real and imaginary components of the out-of-plane scattered wave. (c) In the unperturbed structure, both the real and imaginary components are bound. (d-e) When the top (bottom) row of circles is perturbed into ellipses (black dashed lines), the real (imaginary) part of the quasi-bound wave is coupled to free space but the imaginary (real) part is not. (f,g) FBZ and TM band diagram of the unperturbed structure. (h,i) FBZ and TM band diagram of the perturbed structure, supporting a zone-folded Dirac point. Modes marked by the arrows in (i) correspond to the those in (h), and red and blue crosses marked in (h) correspond to those in (g).

The dual-perturbation scheme sketched in **Figs. 2(d,e)** enables independent control of the real and imaginary parts of the scattered light, which together confer complete command over the surface emission: $(A, \Phi, \psi, \chi)$. Here we choose the fundamental TM guided mode [depicted in **Fig. 1(b)**], which, once coupled into the unperturbed subwavelength lattice, is decomposed into its real and imaginary components [**Fig. 2(c)**]. Each of these components of the travelling TM wave (characterized in the $y$ direction by $e^{-iky}$) is a standing wave of either even or odd parity in the $y$ direction (i.e., cosine or sine). These standing waves abide by selection rules for excitation (or scattering) near the device normal, determining which polarization (if any) couples to free space due to the symmetries broken by the perturbation [38]. The real component is bound except in the presence of Perturbation 1, where the top pair of circles are perturbed into ellipses oriented 90° relative to one another [denoted by the dashed boundaries in **Fig. 2(d)**]. However, Perturbation 1 does not affect the imaginary component, which is symmetry-protected due to its opposite parity. Perturbation 2 has exactly the opposite effect for the same reason: the imaginary component is scattered while the real component is not [**Fig. 2(e)**].

The behavior of a leaky-wave meta-unit can be modeled analytically in combination with full-wave simulations, as described in the **Methods**. **Figure 3(a)** shows two geometric degrees of freedom, $\delta_1$ and $\delta_2$, which determine the sign and strength of each perturbation and hence the signed magnitude of the real and imaginary components of the scattered light. **Figures 3(b,c)** show the amplitude and phase of the scattered light, which is $y$-polarized in this case. At the origin ($\delta_1 = \delta_2 = 0$), a singularity is observed in the phase, corresponding to a null in scattering amplitude, i.e., a bound wave due to the absence of perturbation. This topological feature is a manifestation of the polarization-agnostic geometric phase recently demonstrated to control Fano resonances in nonlocal metasurfaces [46]. Here, we leverage this principle to enable LWMs with complete phase and amplitude (PA) control of any polarization. To produce scattered light with other polarization states, the orientation angles $\alpha_1$ and $\alpha_2$ of the ellipses may be varied [**Fig. 3(d)**]. **Figures 3(e,f)** show the elliptical parameters, $\psi$ and $\chi$, of the scattered light as a function of $\alpha_1$ and $\alpha_2$, with example polarization states drawn for reference; between the dashed contours, arbitrary elliptical states are possible. Collectively, by varying the geometric parameters $(\delta_1, \delta_2, \alpha_1, \alpha_2)$, we can arbitrarily specify the scattered state $(A, \Phi, \psi, \chi)$. The mapping between these parameter spaces, including fine adjustments based on full-wave simulations, are discussed in the **Methods**. As a result, we define a semi-analytical library of meta-units for use in populating a LWM that, upon excitation with a guided wave, produces free-space radiation with desired spatial profiles of amplitude, phase, and polarization.

Finally, the amplitude and phase distributions of the guided portion of the quasi-bound wave must be accounted for when populating a LWM with meta-units targeting a specific device function (see **Methods** for details). For instance, **Fig. 4(a)** shows a target PA profile producing a focused beam, while **Fig. 4(b)** shows the mode-corrected PA profile taking into account the amplitude and phase evolution associated with the guided mode depicted in **Fig. 4(c)**. Hence, targeting $y$-polarized light, **Figs. 4(d,e)** show the resulting profiles of $\delta_1$ and $\delta_2$. The LWM design was then fabricated using electron-beam lithography and characterized in the near-infrared (see **Methods**). An example photo and a scanning electron micrograph of the fabricated devices are shown in **Figs. 4(f,g)**.

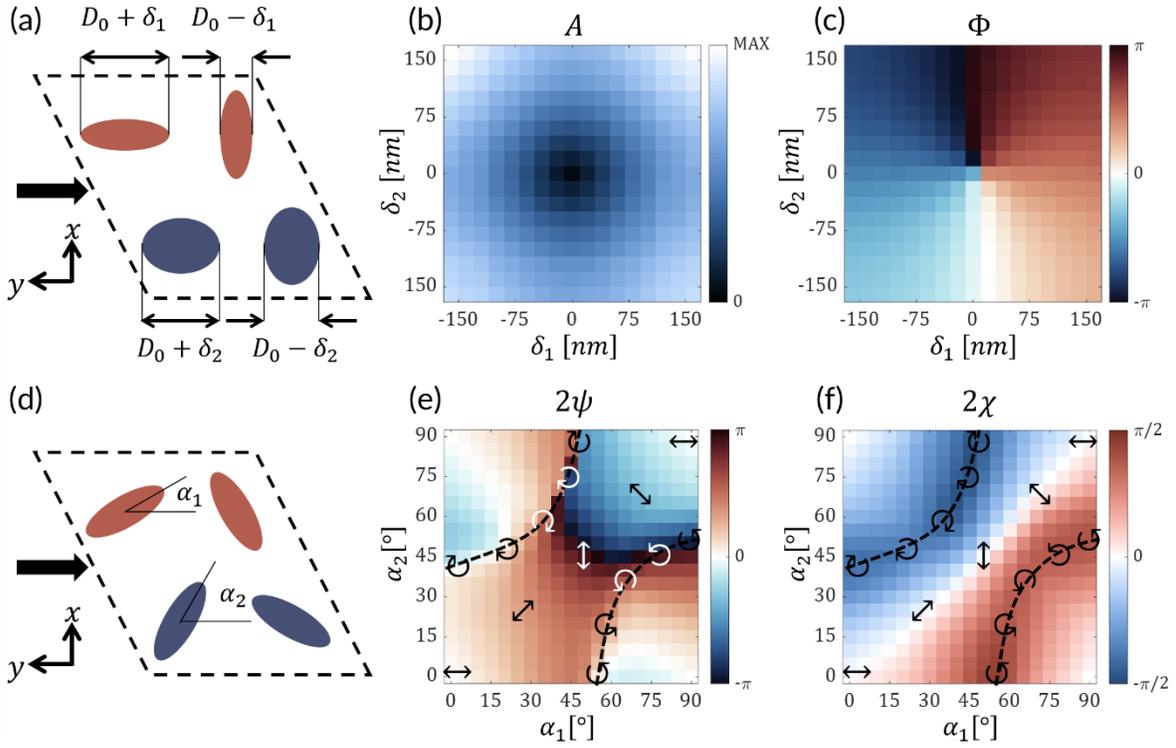

**Figure 3.** Full-wave simulations constructing the meta-unit library. (a) For fixed elliptical orientations, the perturbations $\delta_1$ and $\delta_2$ determine the scattering magnitude of the real and imaginary parts of the field, respectively. (b) Map of scattered amplitude of *y*-polarized light as a function of $\delta_1$ and $\delta_2$, showing a bound state when both perturbations vanish. (c) Map of scattered phase of *y*-polarized light as a function of $\delta_1$ and $\delta_2$, supporting a topological feature characteristic of a geometric phase. (d) For fixed $\delta_1$ and $\delta_2$, the perturbation angles $\alpha_1$ and $\alpha_2$ determine the polarization state scattered by the unit cell. (e,f) Map of $2\psi$ and $2\chi$ as a function of $\alpha_1$ and $\alpha_2$, with dashed contours denoting chiral states near the poles of the Poincaré sphere. Arrows denoting the approximate polarization states in each region are overlaid for reference.

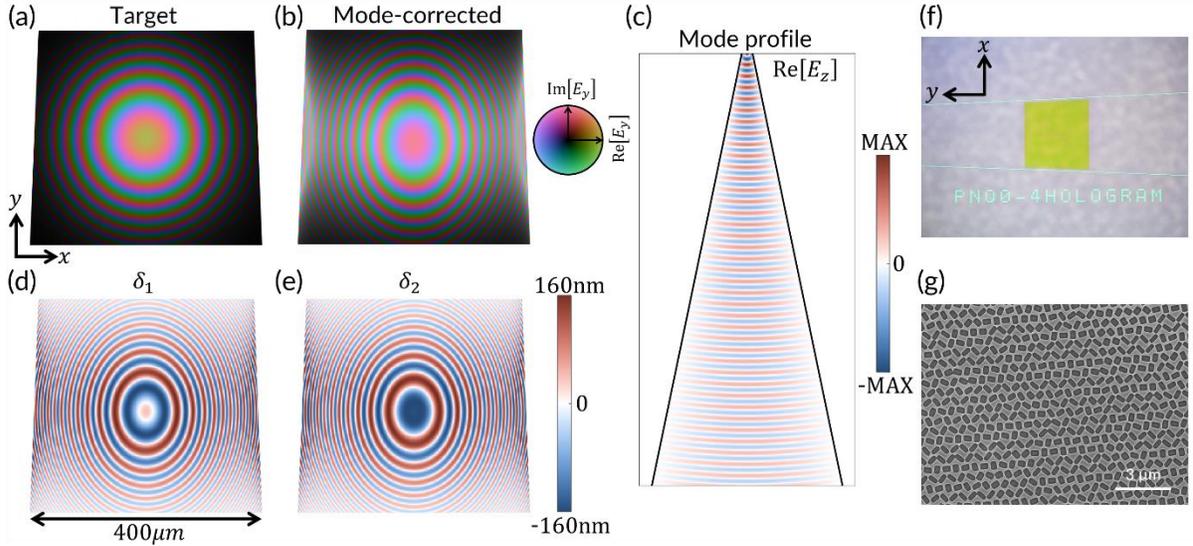

**Figure 4**. Constructing a LWM via modal correction. (a) Target amplitude and phase profiles to produce a converging, linearly polarized beam ($y$-polarization). (b) Mode-corrected amplitude and phase profiles accounting for the guided mode amplitude and phase profiles in (c). (d,e) Spatial profiles of the perturbation strengths $\delta_1$ and $\delta_2$, populated based on the map in (b) from the meta-unit library. (f) Optical micrograph of a fabricated device. (g) Scanning electron micrograph of a fabricated device.

## Phase and amplitude control

We experimentally demonstrate the ability of our LWM platform to generate custom PA wavefronts. We choose meta-unit motifs with fixed angles $\alpha_1$ and $\alpha_2$ so that the wavefronts are linearly polarized. **Figure 5** shows four example LWMs, demonstrating focusing, generation of orbital angular momentum (OAM), PA holography, and a Kagome lattice generator.

First, **Fig. 5(a)** schematically shows a LWM generating a converging beam in the surface-normal direction. As seen in **Fig. 4(a)**, a Gaussian envelope is applied to the device amplitude profile, and the phase profile of a metalens is encoded to focus light at a target focal length $f = 2\ mm$. Longitudinal cross-sections of the measured converging beam are shown in **Figs. 5(b,c)**, at $\lambda = 1530\ nm$. A transverse cross-section at the designed focal plane is shown in **Fig. 5(d)**, where a focal spot is observed with full-widths at half-maximum (FWHM) $w_y = 9.4\ \mu m$ in the $y$ direction and $w_x = 10\ \mu m$ in the $x$ direction. These values are in good agreement with diffraction-limited operation, with simulated values $w_y = 9.4\ \mu m$ and $w_x = 9.1\ \mu m$ [insets of **Fig. 5(d)**]. Images of the focal plane at various operating wavelengths from 1520 $nm$ to 1580 $nm$ are shown in **Fig. 5(e)**. The position of the focal spot along the $y$ direction shifts linearly with respect to the wavelength, following the dispersion of the band diagram in **Fig. 2(i)**, with a dispersion $\frac{d\theta}{d\lambda} = 1.2 \times 10^{-3}\ rad/nm$. Measurements confirming the linearly polarized radiation of this device are shown in **Supplementary Materials S.3**.

Next, **Fig. 5(f)** schematically shows a LWM generating a vortex beam with OAM order $\ell = 2$, in tandem with a tilted wave with a Gaussian profile that serves as an interferometric reference beam [encoded in the complex near field shown in **Fig. 5(g)**]. An image taken at $z = 7\ mm$ shows the interference of the two beams [**Fig. 5(h)**], where a characteristic fork pattern with two branches is formed (confirming the OAM order), while an image taken at $z = 15\ mm$ shows the separation of the two beams [**Fig. 5(i)**]. As another example, **Fig. 5(j)** demonstrates a two-image holographic LWM encoded by the two degrees of freedom inherent to a PA metasurface [complex near field shown in **Fig. 5(k)**]. A first image, the CUNY logo is applied as the amplitude profile of the hologram, while a second image, the Columbia Engineering logo, is encoded in the phase profile of the hologram (using the Gerchberg-Saxton algorithm [47]) such that the logo is reconstructed at a distance of $z = 1\ mm$ (an effective numerical aperture of $NA \approx 0.2$). Images taken at the LWM plane ($z = 0\ mm$) and the holographic image plane ($z = 1\ mm$) are shown in **Fig. 5(l)** and **Fig. 5(m)**, respectively.

As a final example, **Fig. 5(n)** depicts a LWM producing a Kagome lattice via the complex near field distribution shown in **Fig. 5(o)** [**Fig. 5(p)** shows the central region of this distribution]. The measured result at a plane $z = 0.5 \, mm$ away from the LWM (an effective $NA \approx 0.37$) is shown in **Fig. 5(q)**. Additional devices are reported in the **Supplementary Materials S.4**. The detailed near field and geometry profiles of each device in this section and the following section are reported in the **Supplementary Materials S.5**. The evolution of the two-image hologram is shown in **Supplementary Materials S.6**.

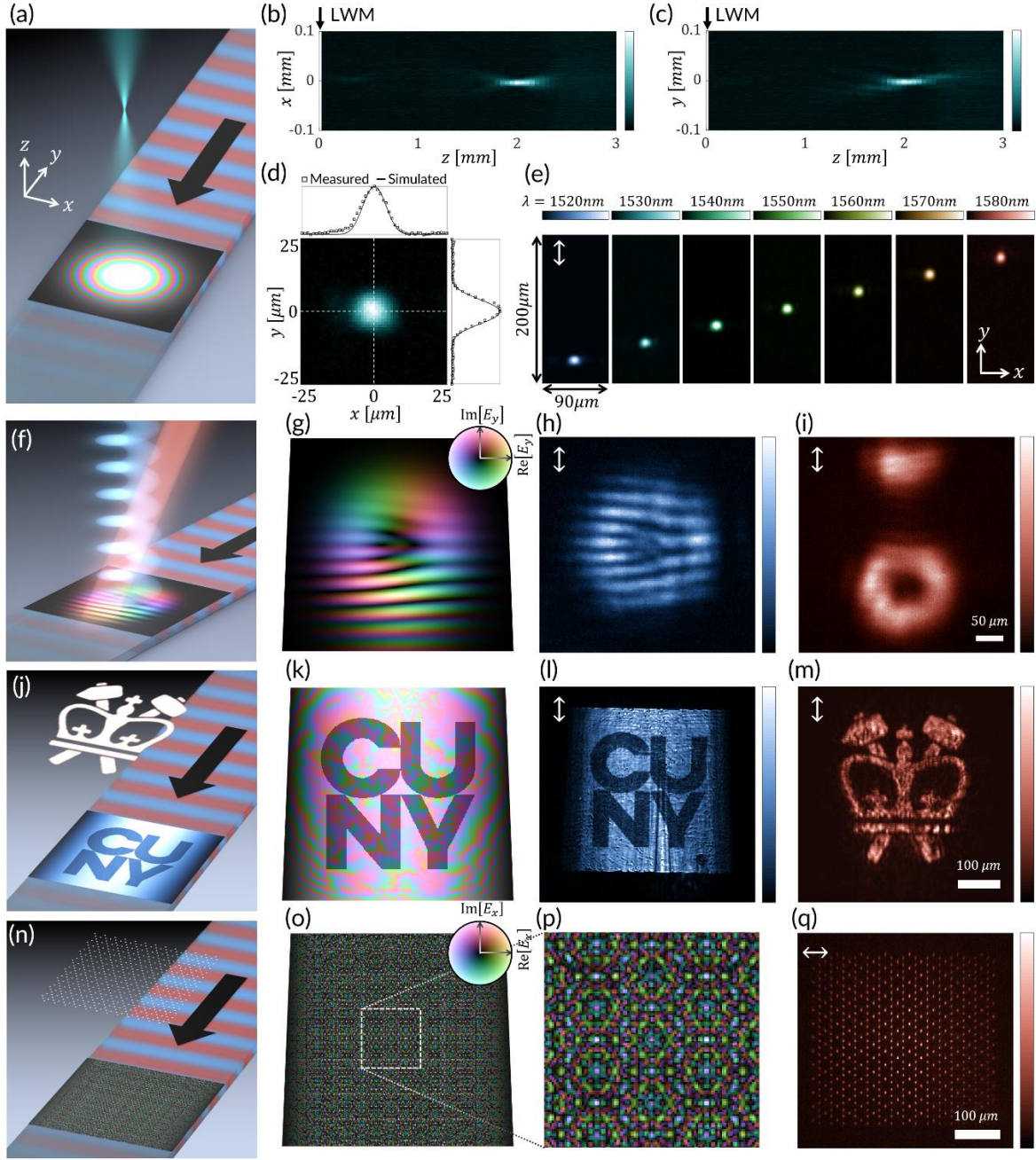

**Figure 5.** Phase-amplitude LWMs for linearly polarized light. (a) Schematic of a focusing LWM. (b,c) Measured $xz$ and $yz$ cross sections showing focused emission from the LWM at $\lambda_0 = 1550\ nm$, with a designed focal length of $f = 2\ mm$. (d) Measured $xy$ cross section at the focal plane for $\lambda_0 = 1530\ nm$, with $x$ and $y$ linecuts compared to simulated responses based on diffraction limited behavior. (e) Measurement of the focal plane at seven selected wavelengths, demonstrating steering in the $y$ direction, following the leaky wave dispersion. (f) Schematic of a LWM producing an OAM beam with $\ell = 2$, along with a tilted Gaussian beam as a reference, via the complex near field in (g). (h) Measured interference of the two beams at a plane $z = 7\ mm$, showing a characteristic forked pattern. (i) Measured emission of the OAM device at a plane $z = 15\ mm$ away from the LWM, where the OAM and Gaussian

beams are separated. (j) Schematic of a two-image hologram, wherein a gray-scale amplitude distribution at the LWM plane serves as a first image, and a distinct holographic image is produced at a second plane based on the phase profile, collectively encoded in the complex near field in (k). (l) Measured gray-scale image (CUNY logo) at the LWM plane. (m) Measured holographic image (Columbia Engineering Logo) at a plane $z = 1\ mm$ away from the LWM. (n) Schematic of a Kagome lattice generator based on the complex near field in (o); the central region of this field is shown in (p). (q) Measured holographic lattice at a plane $z = 0.5\ mm$ away from the LWM. All devices operate for $y$-polarized light, except for the Kagome lattice generator, which produces $x$-polarized light.

**Vector-beam generation**

We next demonstrate LWMs generating vectorial fields. Here, all four geometric degrees of freedom $(\delta_1, \delta_2, \alpha_1, \alpha_2)$ are utilized to realize PA control for the two orthogonal polarization components simultaneously, or the PA profile of a vector beam. **Figures 6(a,b)** schematically show a four-image holographic LWM, extending the scheme in **Figs. 5(j-m)**. Images of the letters "$\psi$" and "$\chi$" are applied to the amplitude profiles of the $y$ and $x$ polarization components of the scattered field, while the phase profiles encode the letters "$A$" and "$\Phi$" respectively, for reconstruction at a distance of $z = 1\ mm$ [**Figs. 6(a,b)**]. Images taken at the holographic image plane ($z = 1\ mm$) and the LWM plane ($z = 0\ mm$) for $y$-polarization are shown in **Fig. 6(c)** and **Fig. 6(d)**, respectively; **Figs. 6(e,f)** depict the same for $x$-polarization.

Finally, **Fig. 6(g)** schematically shows a LWM generating a focused Poincaré beam with minimum waist size at a distance of $z = 2\ mm$. Here, we implement the Poincaré beam as the superposition of a focused left-circularly-polarized (LCP) Gaussian beam and a focused right-circularly-polarized (RCP) vortex beam with $\ell = 1$, so that a transverse cross-section of the beam reveals all polarization states over the Poincaré sphere [**Fig. 6(g)**]. The four optical degrees of freedom we control in this specific demonstration are the amplitude and phase profiles of the LCP and RCP states. **Figure 6(h)** shows the measured intensity distributions at a distance $z = 2\ mm$ and at six characteristic polarization states, in good agreement with the simulated results in **Fig. 6(i)**. The evolution of the four-image hologram is shown in **Supplementary Materials S.6**.

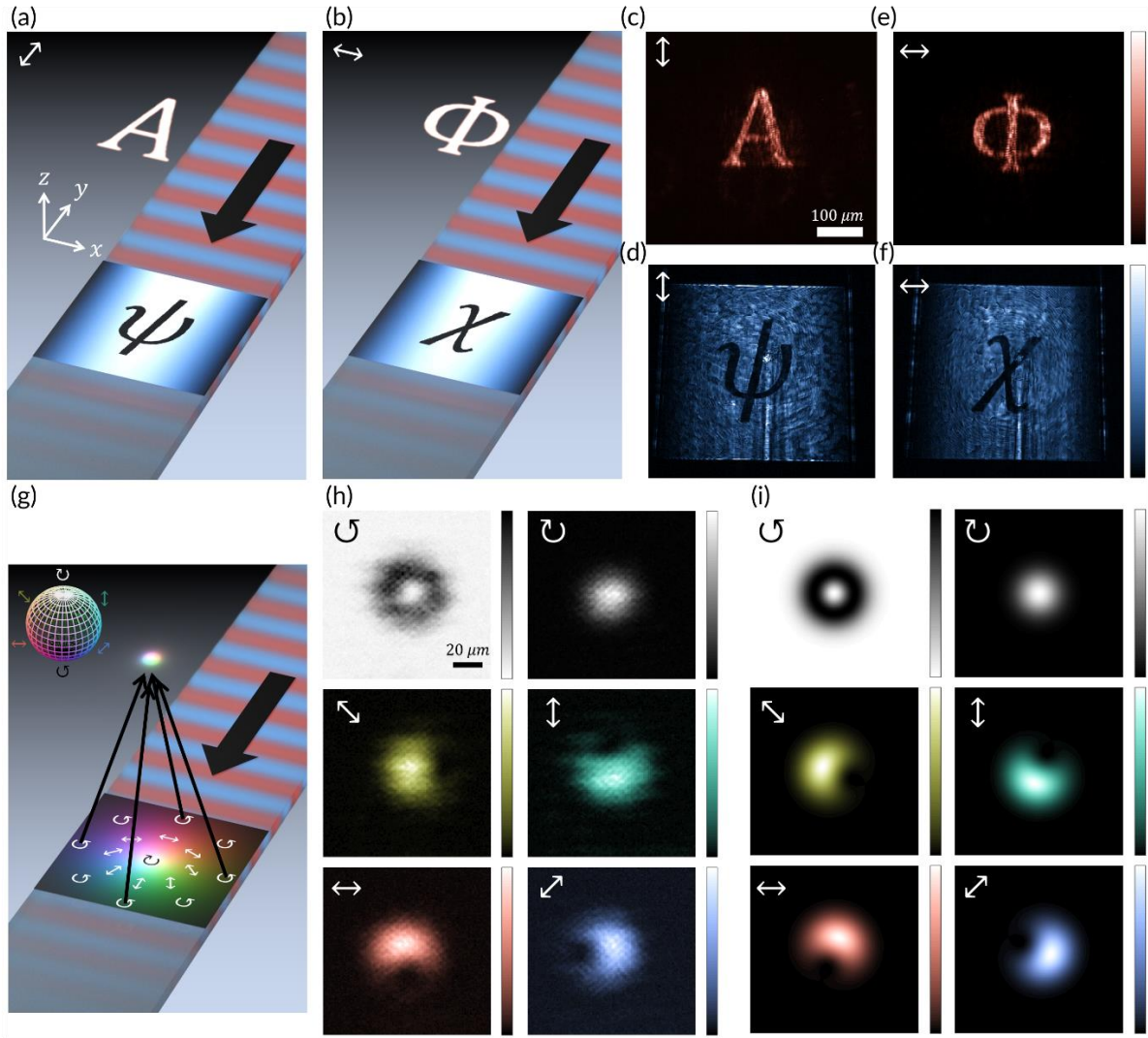

**Figure 6.** Vector-beam LWMs with complete control over amplitude, phase, and polarization. (a,b) Schematic of a LWM producing a four-image hologram, in which a two-image hologram is encoded simultaneously for $x$ and $y$ polarizations. (c,d) Measured holographic image ($z = 1\ mm$) and LWM ($z = 0\ mm$) planes imaged through a polarizer allowing $y$-polarized light to pass. (e,f) Measured holographic image ($z = 1\ mm$) and LWM ($z = 0\ mm$) planes imaged through a polarizer allowing $x$-polarized light to pass. (g) Schematic of a LWM producing a focusing Poincaré beam. Measured (h) and simulated (i) profiles of six characteristic polarizations at a plane $z = 2\ mm$.

**Outlook and conclusions**

In this work, we demonstrated complete command over leaky radiation from LWMs through a rational design approach based on quasi-bound states in the continuum that originate from broken symmetries. Our approach confers a number of novelties and advantages compared to other techniques. The novelties center around the period-doubling perturbation, which exclusively introduces coupling to free-space—the mode is otherwise bound. This feature in principle allows for arbitrarily large aperture fields, and here we demonstrated surface emission from integrated devices with a linear dimension $> 250\lambda_0$. The meta-unit motif with two rows of apertures shifted from each other enables simultaneous and independent control of both amplitude and polarization state of both real and imaginary components of the LWM radiation. The design's underlying origins in symmetry considerations enables a semi-analytical mapping of the optical degrees of freedom $(A, \Phi, \psi, \chi)$ to and from the geometric design parameters $(\delta_1, \delta_2, \alpha_1, \alpha_2)$. Traditionally, the difficulty of constructing a meta-unit library compounds unfavorably as the number of targeted optical degrees of freedom is increased. Here, in contrast, the LWM geometry is populated point-by-point based on a set of simple equations, while achieving complete control over the vectorial field. Further, the lattice supports a zone-folded Dirac cone, enabling operation at and near the device normal (broadside emission), a feature precluded by the parabolic band structure of modes employed in conventional GC designs.

    Our symmetry-based design principle implies that applications involving a wide array of materials and frequencies may adopt this approach. For instance, RF leaky-wave antennas are well-known for beamforming and scanning in the far-field, but are difficult to operate at close range. Our approach may be used to create RF leaky-wave antennas that

operate at a wide range of distances, and with distinct functionalities imparted to orthogonal polarizations, useful for polarization-division multiplexing. Similarly, in the context of PICs, while we showed here one popular material platform based on silicon nitride, the design principle can be applied to silicon-on-insulator technologies. Notably, active materials such as lithium niobate, 2D materials and liquid crystals may also be incorporated, in order to switch on or off the symmetry-breaking perturbation or to control its magnitude. Finally, while our LWMs are composed of holes in a highly symmetric array, subwavelength grating waveguides [48] composed of pillars may also be used based on the same principles.

Several extensions may also be explored. First, two-layer devices based on similar period-doubling symmetry-breaking principles have shown exquisite control over the leakage of chiral states [40]. Here, our approach has been primarily achiral: due to the insignificant breaking of out-of-plane symmetry, the upward and downward radiating states are mirror images of each other. In contrast, two-layer devices may add additional control to manipulate separately the upward and downward radiation. Similarly, multi-perturbation devices based on symmetry-breaking have shown control over several leaky waves simultaneously [39],[49]. Here, our approach controlled a single mode, but future work may extend our platform to control orthogonally propagating modes. Next, while here we used a weakly corrugated system without deliberate command of the group velocity, band structure engineering (see, e.g., [50],[51]) may be used to tune the angular dispersion of the output. Last, while here we implemented a single device layer, due to the broadband transparency of these nonlocal metasurfaces to free-space light, future works may cascade several LWMs at optically thick distances for multi-wavelength operation [49].

Finally, we highlight a few improvements that may be explored to further extend the impact of this work. First, while we showed compatibility with large aperture fields, we made no attempt to optimize the scattering for larger radiation efficiency. Future efforts may explore matching the aperture radiation field with the amplitude profile of the quasi-bound wave to fully scatter the incident guided wave [52]. Second, while in our case the flat band observed in **Fig. 2(i)** did not negatively impact the function of our device, in deeply corrugated structures scattering between the Dirac cone and this flat band may introduce unwanted cross-talk. This may be avoided by adjusting the lattice dimensions $a_x$ and $a_y$. Last, here we limited our operation to near-normal surface emission (with moderately high effective numerical apertures in the range of $NA \approx 0.1 - 0.37$), allowing us to decouple the real and imaginary components of the scattered wave via their distinct symmetries. At large $NA$ operation or at extreme deflection angles, this assumption may be invalid (to varying degrees in different systems), implying that more complex meta-unit design must be taken into account.

In conclusion, we have introduced a leaky-wave metasurface platform that generates custom vectorial field at will, combining the functionality of metasurfaces with the compact form-factor of GCs. The design principles are rooted in the symmetries of quasi-bound waves supported by high-symmetry lattices, and as such are compatible with a wide range of material platforms and frequencies. We demonstrated semi-analytical generation of a library of meta-units with complete command over amplitude, phase, and polarization state of light with subwavelength resolution across large aperture fields ($> 250\lambda_0$). In the future, we anticipate a number of applications stemming from this approach. Notably, our platform may be integrated with a PIC for off-chip communications such as chip-to-chip communications and free-space mode-division multiplexing [e.g., using OAM, **Figs. 5(f-i)**, or Poincaré beams, **Figs. 6(g-i)**], and it may be used to generate custom cold-atom traps for quantum applications [such as the Kagome lattice in **Figs. 5(n-q)**]. Our approach may also enable LIDAR systems with arbitrary beamforming (including broadside emission) originating from optically large apertures, tunable with wavelength [**Fig. 5(e)**]. Finally, while our implementation employs structural birefringence as a perturbation, small changes in material birefringence (such as in liquid crystals) may achieve similar control but in a dynamic manner. In this way, our work paves the way towards novel holographic display technologies [**Figs. 5(j-m)** and **Figs. 6(a-f)**], quantum photonic devices, and next-generation communications and sensing.

## Methods

### Modeling and simulations of the meta-units

The Jones vector response of a half meta-unit composed of one pair of ellipses, and excited by a TM slab waveguide mode, approximately follows

$$\begin{pmatrix} E_x \\ E_y \end{pmatrix} = \delta \begin{pmatrix} a_x \sin 2\alpha \\ a_y \cos 2\alpha \end{pmatrix},$$

where $a_x$ ($a_y$) is the maximum amplitude in $x$ ($y$) polarization, which is achieved at maximum $\delta$ and $\alpha = 45°$ ($0°$). The imbalance between $a_x$ and $a_y$ originates from the asymmetry between x and y directions of the system. Here only the first-order perturbation effect is considered, in which regime the dependence on $\delta$ is approximately linear. If the meta-unit is excited by a transverse electric (TE) slab waveguide mode instead, the Jones vector response will be rotated 90° due to the conversion between E-field and H-field, with an additional constant term in $E_x$ from zeroth-order scattering.

The response of a half meta-unit covers an ellipse-disk range, where the amplitude, $A$, and the polarization orientation angle, $\psi$, can be independently controlled by $\delta$ and $\alpha$, respectively. For a given target $A$ and $\psi$, the corresponding meta-atom parameters can be explicitly solved:

$$\alpha = \frac{1}{2}\text{arccot}\left(\frac{a_x}{a_y}\tan\psi\right),$$

$$\delta = \frac{A}{a_x}\cos\psi\sqrt{1+\left(\frac{a_x}{a_y}\tan\psi\right)^2}.$$

For a full meta-unit composed of two pairs of *p2* meta-atoms displaced by a quarter period along the propagation direction, the response is a coherent summation from the two meta-atom pairs with a 90° phase difference:

$$\begin{pmatrix}E_x\\E_y\end{pmatrix} = \delta_1\begin{pmatrix}a_x\sin 2\alpha_1\\a_y\cos 2\alpha_1\end{pmatrix} + i\delta_2\begin{pmatrix}a_x\sin 2\alpha_2\\a_y\cos 2\alpha_2\end{pmatrix},$$

where the first pair of meta-atoms $(\delta_1, \alpha_1)$ contributes to the real part, and the second pair of meta-atoms $(\delta_2, \alpha_2)$ contributes to the imaginary part. For a complex target $(E_x, E_y)^T$, the corresponding meta-atom parameters can be solved by first calculating the amplitudes $A_{1,2}$ and polarization orientation angles $\psi_{1,2}$ of the real and imaginary parts, respectively:

$$A_1 = \sqrt{[\text{Re}(E_x)]^2 + [\text{Re}(E_y)]^2}, \quad A_2 = \sqrt{[\text{Im}(E_x)]^2 + [\text{Im}(E_y)]^2},$$

$$\psi_1 = \arctan\frac{\text{Re}(E_y)}{\text{Re}(E_x)}, \quad \psi_2 = \arctan\frac{\text{Im}(E_y)}{\text{Im}(E_x)},$$

then using the above formulas for each half meta-unit:

$$\alpha_{1,2} = \frac{1}{2}\text{arccot}\left(\frac{a_x}{a_y}\tan\psi_{1,2}\right),$$

$$\delta_{1,2} = \frac{A_{1,2}}{a_x}\cos\psi_{1,2}\sqrt{1+\left(\frac{a_x}{a_y}\tan\psi_{1,2}\right)^2}.$$

FDTD simulation results of the polarization ellipticity angle ($\chi$) and the polarization orientation angle ($\psi$) responses of a leaky-wave meta-unit are shown in Figs. 3(e,f), where $\alpha_1$ and $\alpha_2$ are swept from 0° to 90° to cover the polarization space once, and $\delta_{1,2}$ is set according to $\alpha_{1,2}$ to generate a flat amplitude response. A full coverage of polarization ellipticity angle from $\chi = +\pi/4$ (LCP) through $\chi = 0$ (LP) to $\chi = -\pi/4$ (RCP) is achieved as $\alpha_1 - \alpha_2$ varies from +45° through 0° to −45° [Fig. 2(f)], and a full coverage of polarization orientation angle from $\psi = -\pi/2$ to $\psi = \pi/2$ is achieved as $\alpha_1$ and $\alpha_2$ varies from 0° to 90° in the $|\alpha_1 - \alpha_2| < 45°$ region [Fig. 3(e)]. Further simulation results of the amplitude ($A$) and the phase ($\Phi$) responses are shown in Fig. 3(b) and (c), respectively, where $\alpha_1 = \alpha_2 = 0°$ is set to fix the polarization at $\chi = \psi = 0$ (y-polarized). Full, symmetric coverage of amplitude and phase is achieved along the radial [Fig. 2(b)] and azimuthal [Fig. 3(c)] directions on the parameter sweep, respectively.

**Device fabrication**

We experimentally demonstrate leaky-wave metasurfaces at the telecommunications wavelengths ($\lambda \approx 1.55$ μm) using a polymer-$Si_3N_4$ materials platform. Both the waveguide circuit and the meta-antenna holes composing the metasurface are patterned in a 300-nm polymer (PMMA) layer, on the top of a 300-nm $Si_3N_4$ planar layer, on a $SiO_2$ substrate. The fundamental TM guided mode is fed from a single-mode ridge waveguide via a linear taper with a taper rate of $\Delta w/\Delta L = 1/12$. The metasurface with a size of ~400 μm is integrated in the taper (**Fig. 2f**). The amplitude and phase distributions in the tapered slab waveguide is nonuniform, which should be compensated for when designing the metasurface profiles. Expanded from a single-mode waveguide through a large linear taper, the slab waveguide source profile of the metasurface region can be approximated with an analytic expression:

$$E(x,y) = E_0 A(x,y) e^{i\phi(x,y)},$$

with

$$A(x,y) = \sqrt{\frac{w_0}{w(y)}} \cos\left[\frac{\pi x}{w(y)}\right] e^{\frac{\alpha y}{2}},$$

$$\phi(x,y) = \frac{2\pi}{\lambda}\left[-n_{ms} y + n_{wg} \frac{x^2}{2(L-y)}\right],$$

where the amplitude distribution $A(x,y)$ is formed by the collective effects of: (i) waveguide widening, i.e., $w(y) = w_0 + \Delta w/\Delta L \cdot (L - y)$, where $L \approx 4800$ μm is the taper length and $y$ is the longitudinal coordinate in the metasurface region, (ii) transverse waveguide mode distribution with a width of $w(y)$, and (iii) radiation attenuation due to the scattering from the metasurface; the phase distribution $\phi(x,y)$ is composed of: (i) longitudinal phase accumulation in the metasurface region with an effective modal index $n_{ms} \approx 1.52$, and (ii) transversal phase accumulation in the linear taper with an effective modal index $n_{wg} \approx 1.55$, approximated by a paraxial cylindrical wave.

Finally, the devices are fabricated as follows. SiN thin films of 300 nm thickness are grown via plasma-enhanced chemical vapor deposition on a fused silica substrate of 180-μm thickness. A 300 nm-thick layer of poly(methyl methacrylate) is spin coated and baked at 180 °C to serve as an electron-beam resist. EBL (Elionix ELS-G100) is then carried out at 100 keV and 1 nA, with a dose of 750 μC/cm2 and appropriate proximity effect corrections (BEAMER) to define the waveguide border and metasurface aperture patterns. A 3:1 mixture of isopropyl alcohol to deionized water is used to develop the exposed resist. The fabricated chip is then cleaved cross the narrow waveguide segment for fiber coupling.

**Device characterization**

Near-infrared light at $\lambda \approx 1.55$ μm is produced by a diode laser, and is coupled into the device using a lensed optical fiber with proper polarization adjustment. The free-space light on the air side generated by the metasurface is collected by a 10× or 20× near-infrared objective (Mitotoyu), passed through a polarization filter (Thorlabs), and directed towards a near-infrared camera (Princeton Instruments).

**Acknowledgements**

This work was supported in part by the Air Force Office of Scientific Research MURI program and the Simons Foundation.

**Author contributions**

A.C.O conceived the idea. H.H. and A.C.O mathematically modeled the devices and performed simulations. H.H. designed and fabricated the devices. H.H., A.C.O., Y.X., S.C.M., and C.-C. T. characterized the devices. A.C.O and H.H. analyzed the data. A.C.O., H.H., N.Y., and A.A. wrote the manuscript. N.Y. and A.A. supervised the research.

**Competing interests**

The authors declare no competing interests.

# Supplementary Materials for

# Leaky-wave metasurfaces for integrated photonics

Heqing Huang[1,†], Adam C. Overvig[1,2,†], Yuan Xu[1], Stephanie C. Malek[1], Cheng-Chia Tsai[1], Andrea Alù[2,3,*], and Nanfang Yu[1,*]

[1]Department of Applied Physics and Applied Mathematics, Columbia University, New York, NY 10027, USA.

[2]Photonics Initiative, Advanced Science Research Center, City University of New York, New York, NY 10031, USA

[3]Physics Program, Graduate Center of the City University of New York, New York, NY 10016, USA

†Authors contributed equally

*Corresponding authors: ny2214@columbia.edu, aalu@gc.cuny.edu

## S.1 Recent progress in metasurfaces on waveguides

By incorporating arrays of optical nano-antennas with waveguides, integrated metasurfaces have been developed for conversion between distinct guided modes and radiation modes, but rarely for generating free-space wavefronts that are extended in two dimensions and with complex profiles. Waveguide-holograms have been proposed long ago as a generalization of grating couplers, but based on simple structural units and were designed to achieve only amplitude or phase control on the free-space wavefront, while the rich degrees of freedom in the meta-units of metasurfaces remain to be explored.

To achieve full control over the four optical degrees of freedom, at least four structural degrees of freedom is needed in a meta-unit. This has recently been realized in free-space metasurfaces, where each meta-unit is composed of four rectangular meta-atoms with variable lengths and widths [Wu *et al. Adv. Optical Mater.* (2022) 10, 2101223] or additional rotation angles [Liu *et al. Light: Science & Applications*. (2021) 10, 107]. However, limited control over the optical degrees of freedom has been realized in integrated metasurfaces. Typically, single phase control [Fang *et al. Nanophotonics* (2022) 11, 1923–1930 and Ding *et al. ACS Photonics* (2022) 9, 398–404] or amplitude control [Huang *et al. Optica* (2019) 6, 119–124] was used to manipulate the wavefront generated by an integrated metasurface. By utilizing two structural degrees of freedom in a meta-unit, integrated metasurfaces have been demonstrated to multiplex free-space wavefronts with respect to incident directions of guided waves [Ha *et al. Adv. Theory Simul.* (2021) 4, 2000239 and Livneh *et al. arXiv*:2112.04152 (2021)]. In this work, we demonstrate for the first time complete independent control over the four optical degrees of freedom using an integrated metasurface with four structural degrees of freedom in a meta-unit.

**Table S.1**: Summary of integrated metasurfaces generating 2D wavefronts

| Work | Structural degrees of freedom | Optical degrees of freedom | |
|---|---|---|---|
| Huang *et al. Optica* (2019) 6, 119–124 | 1 | 1 | $A$ |
| Fang *et al. Nanophotonics* (2022) 11, 1923–1930 | 1 | 1 | $\Phi$ |
| Ding *et al. ACS Photonics* (2022) 9, 398–404 | 2 | 1 | $\Phi$ |
| Ha *et al. Adv. Theory Simul.* (2021) 4, 2000239 | 2 | 2 | $\Phi_1, \Phi_2$ |
| Livneh *et al. arXiv*:2112.04152 (2021) | 2 | 2 | $A_1, A_2$ |
| This work | 4 | 4 | $A, \Phi, \psi, \chi$ |

## S.2 Spectrally shifting the flat band

Here we show that the unwanted flat band overlapping the Dirac cone may be spectrally shifted by manipulating the periodicity of the metasurface lattice in the x direction. **Figure S1(a)** shows a lattice with $a_x = 0.35$ μm and $a_y = 0.4$ μm, which is adjusted compared to the lattice used in the main text [**Fig. S1(b)**], where $a_x = a_y = 0.4$ μm. The lattice may be further adjusted to have a $a_x = 0.45$ μm [**Fig. S1(c)**]. The band diagrams of each case are provided in **Figs. S1(d-f)**, showing that the flat band redshifts with increasing value of $a_x$, and that it is degenerate with the Dirac cone when $a_x = a_y$. The shrunken lattice in **Fig. S1(a)** is a good option for blueshifting the flat band out of the spectral region of interest, avoiding any potential complications entirely.

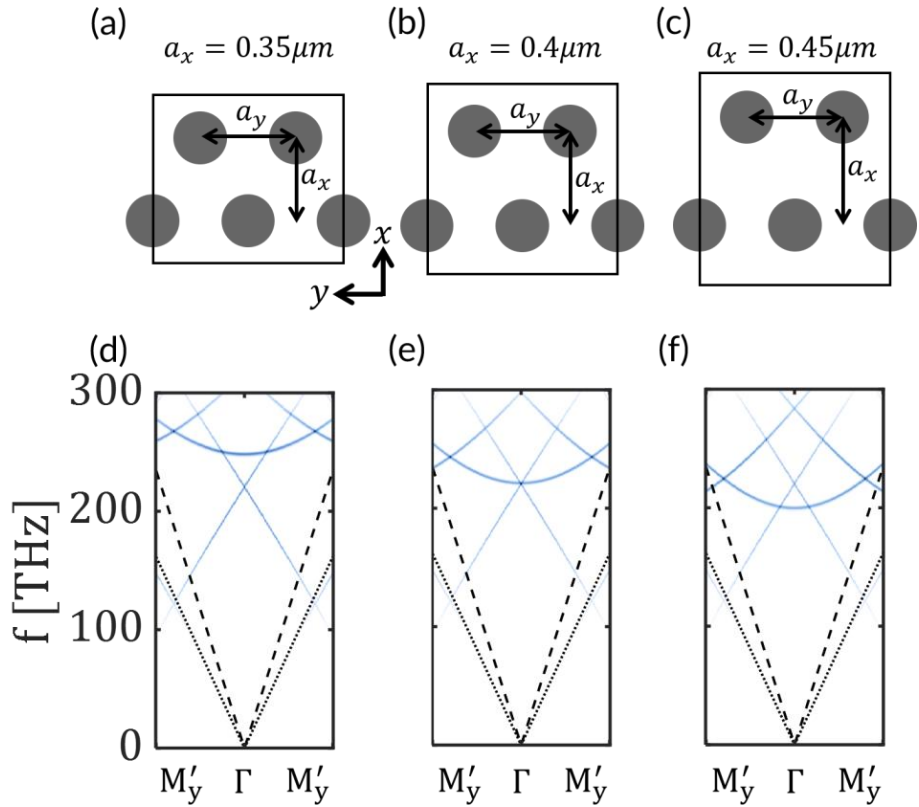

**Figure S1**. (a-c) Lattices with varying values of $a_x$ but fixed value of $a_y = 0.4$ μm. (d-f) The band diagrams of the period-doubled lattices in (a-c), showing that the flat band is tuned by the lattice parameter $a_x$, while the Dirac cone is minimally affected.

## S.3 Polarization-resolved measurement of the focusing LWM

Here we report corroborating experiments demonstrating that the phase-amplitude devices emit linearly polarized light. **Figure S2** depicts measurements using a polarizer to confirm that the emission of the focusing LWM in the main text is linearly polarized along the $y$ direction.

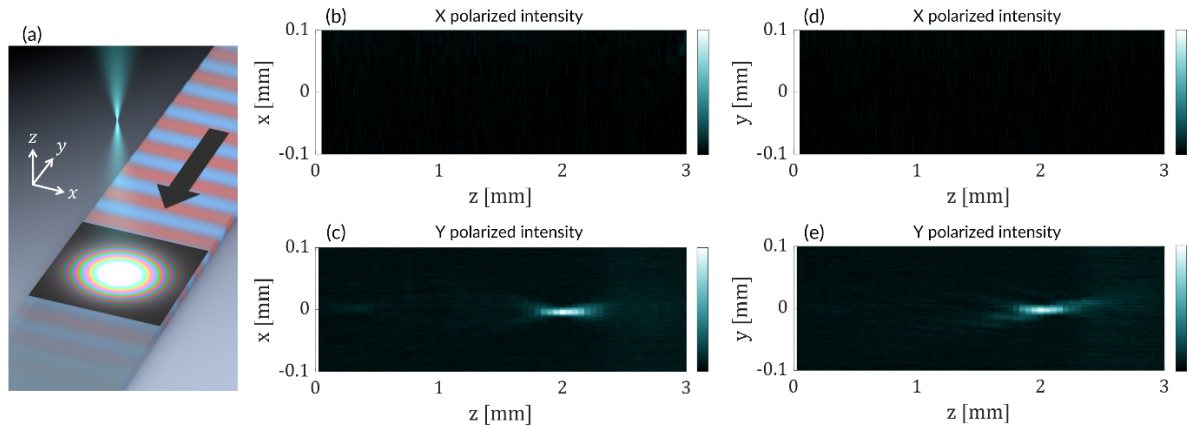

**Figure S2.** (a) Schematic of an LWM generating a focusing beam that is y-polarized. (b) and (c) Measured intensity distributions on the $xz$ plane through an $x$ (b) and $y$ (c) linear polarizer. (d) and (e) Measured intensity distributions on the $yz$ plane through an $x$ (d) and $y$ (e) linear polarizer. The measurements use the same integration times and are plotted on the same color scale.

## S.4 Additional experimental devices

Here we report two additional devices demonstrated experimentally, including a vortex beam generator with orbital angular momentum (OAM) $\ell = 1$ [**Figs. S3(a-c)**] and a phase-only Kagome lattice generator [**Figs. S3(d-e)**]. The former device further evidences the customizability of LWMs as a platform for free-space communications using OAM. The latter device has only phase control over the device plane, utilizing more of the incident power to generate a Kagome lattice, but at the cost of phase control of each point in the lattice (which the PA Kagome in the main text maintains).

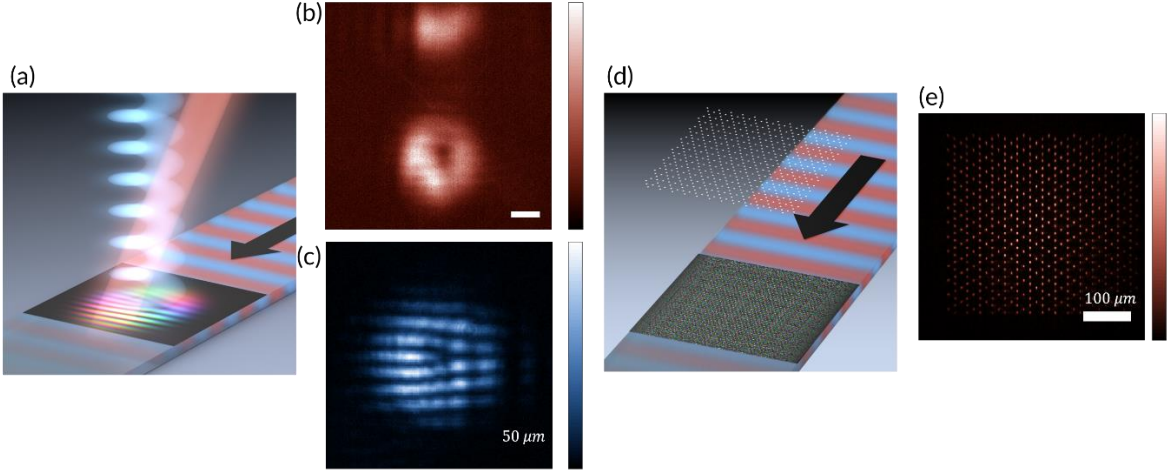

**Figure S3.** (a) Schematic of an LWM producing an OAM beam with $\ell = 1$, along with a tilted Gaussian beam as a reference. (b) Measured emission of the device at a plane $z = 15\ mm$ away from the LWM, where the OAM and Gaussian beams are separated. (c) Measured interference of the two beams at $z = 6.65\ mm$, showing a characteristic forked pattern. (d) Schematic of a Kagome lattice generator with only phase control. (e) Measured Kagome lattice at a plane $z = 0.5\ mm$ away from the LWM.

## S.5 Details of devices shown in the main text

We report the details of the amplitude, phase, and polarization profiles, as well as the geometry profiles, for the devices in the main text. In each box below (black solid lines), we report the target fields (i.e., near-fields immediately above the metasurface, in terms of $(A, \Phi, \psi, \chi)$), mode-corrected field profiles (i.e., fields corrected by the intrinsic amplitude and phase profiles of the waveguide mode, in terms of $(A, \Phi, \psi, \chi)$), and device geometry (i.e., perturbation parameters across the device, in terms of $(\delta_1, \delta_2, \alpha_1, \alpha_2)$). For the linearly polarized devices, the values of the perturbation angles are fixed to be $\alpha_1 = \alpha_2 = 0°$, except for Fig. 5(n), in which case we set $\alpha_1 = \alpha_2 = 45°$. Each box is labeled with the subfigure of the corresponding device schematic in the main text or in **Fig. S3**.

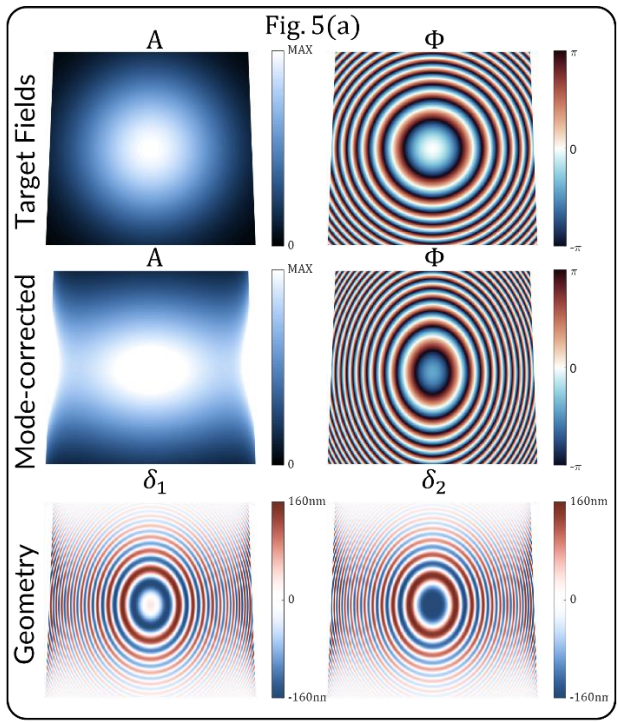
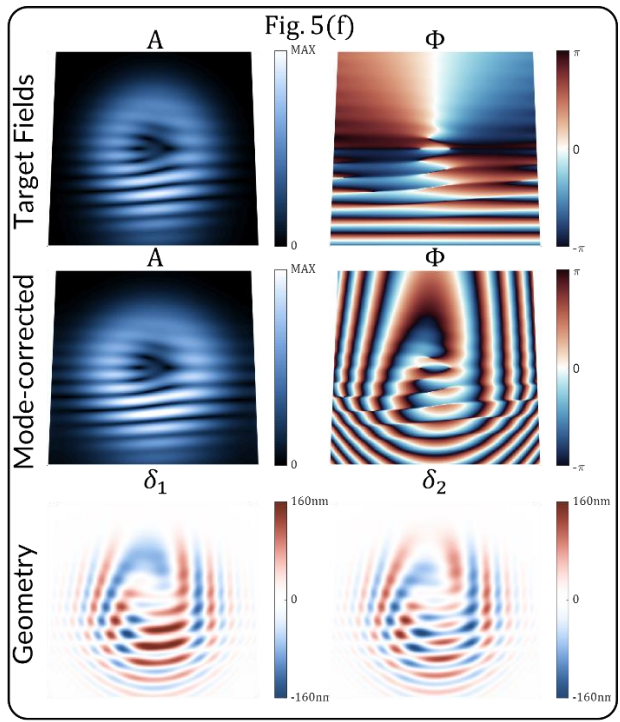
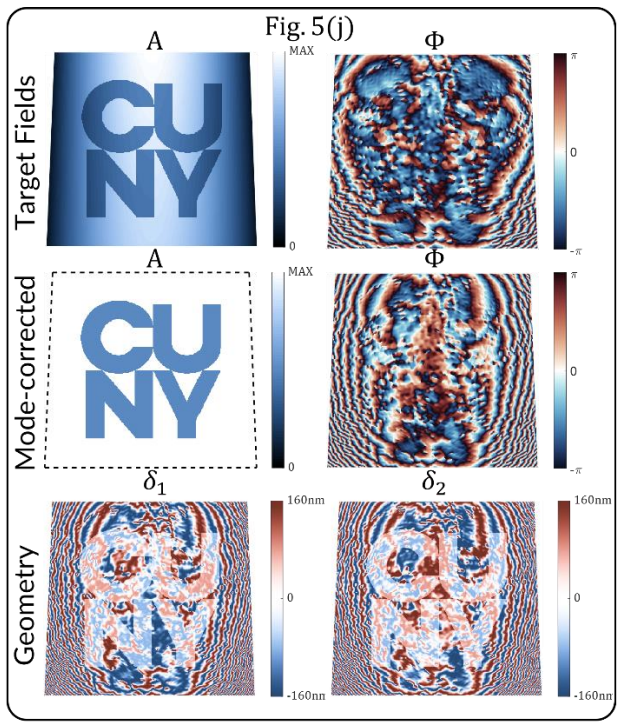
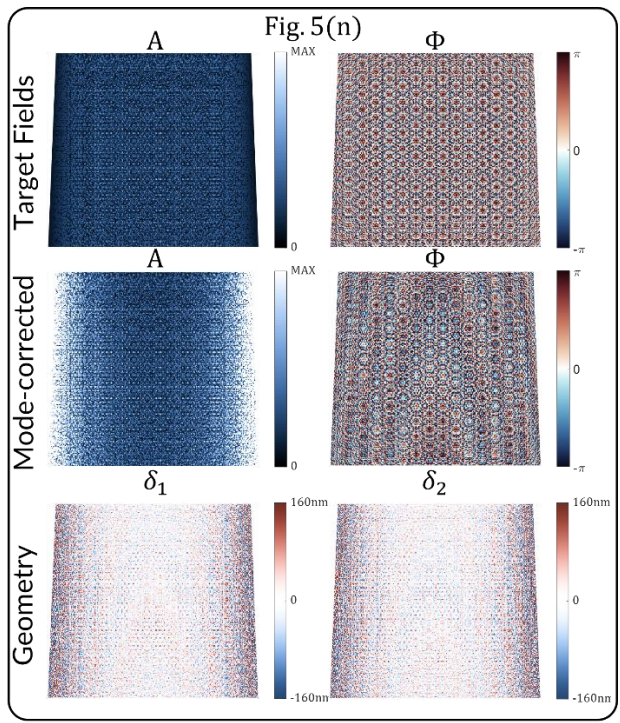

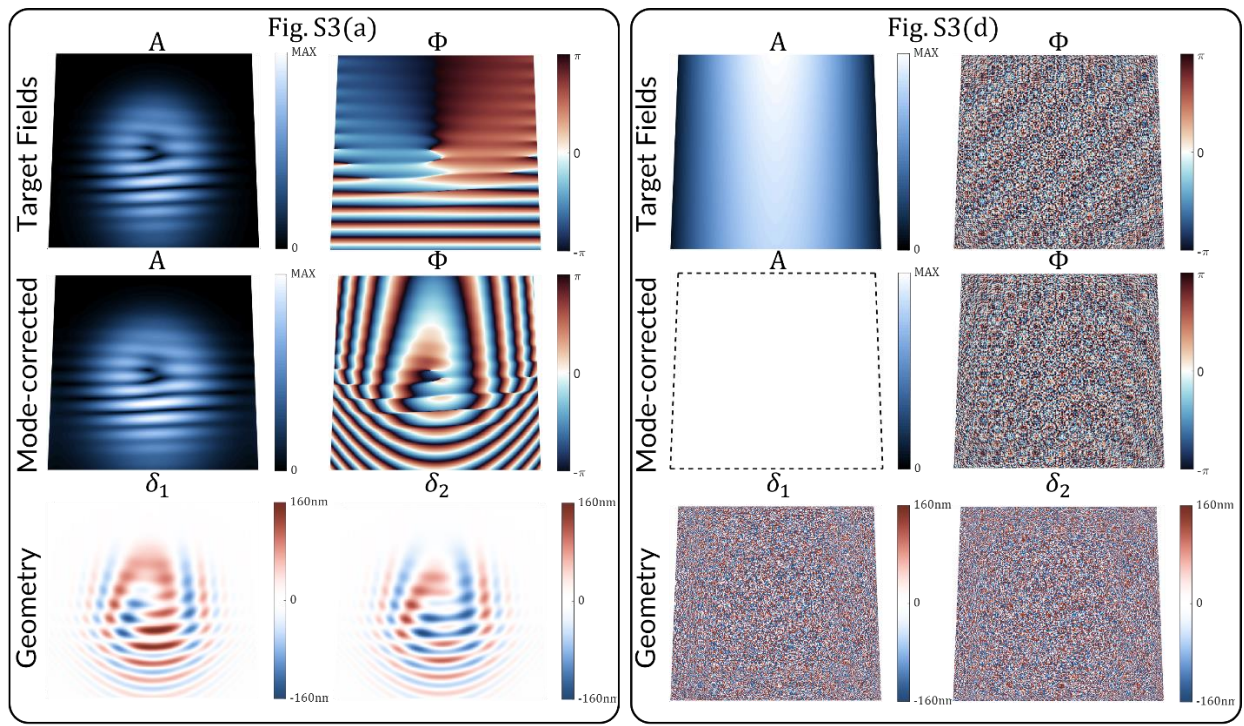

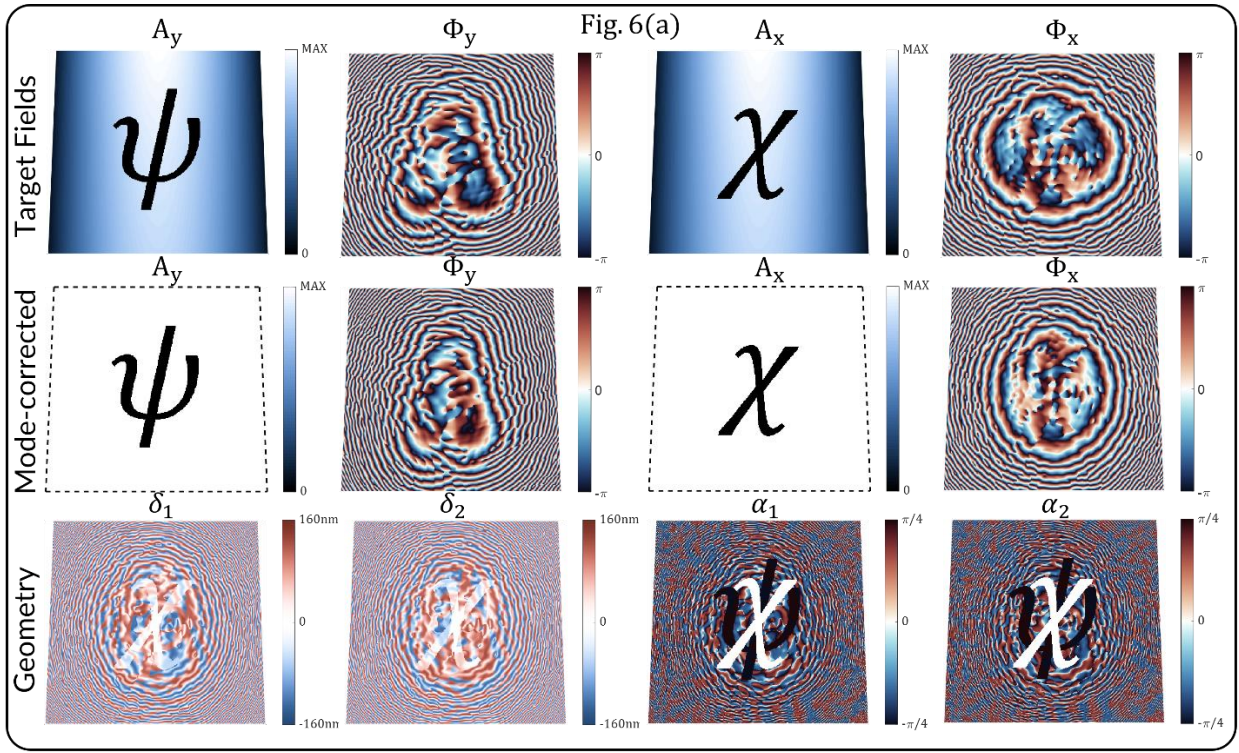

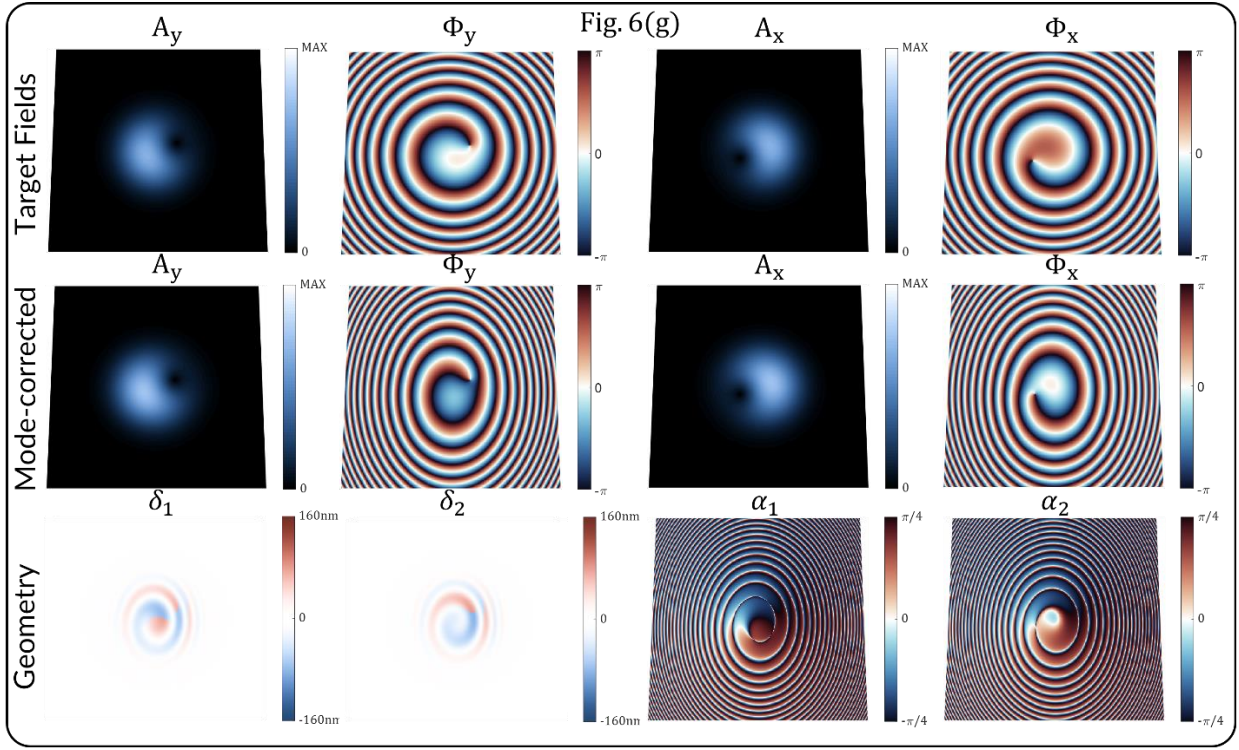

## S.6 Details of devices shown in the main text

**Figure S4** depicts the evolution of transverse optical intensity distributions for the two-image and four-image holograms in the main text, propagating from the LWM plane ($z = 0\ mm$) to the holographic plane ($z = 1\ mm$).

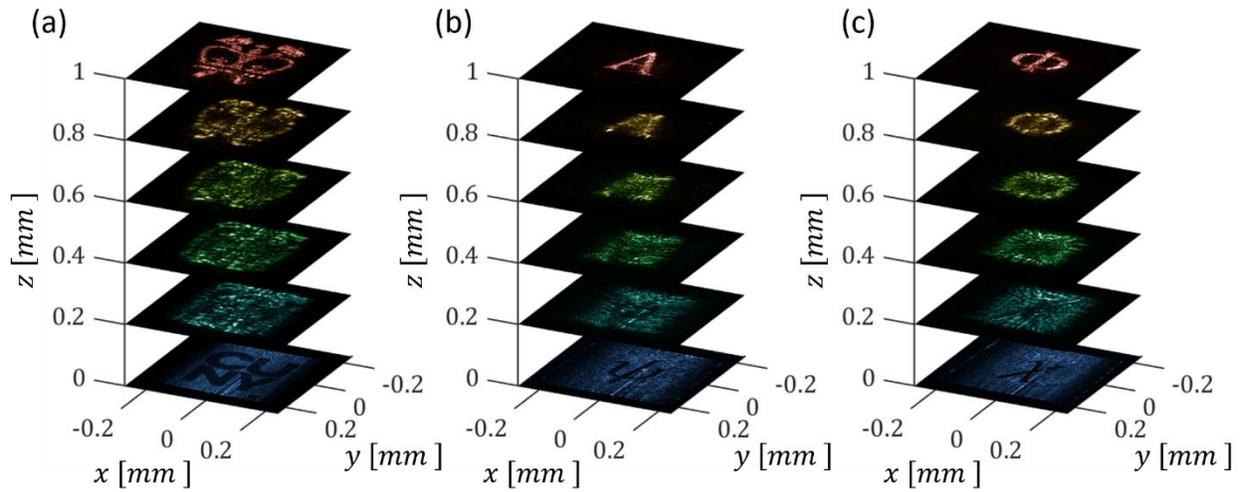

**Figure S4.** (a) Experimental optical intensity distributions of the two-hologram device at six cross-sectional planes. (b) Experimental optical intensity distributions of the four-hologram device at six cross-sectional planes, for $y$ polarized light. (c) Experimental optical intensity distributions of the four-hologram device at six cross-sectional planes, for $x$ polarized light.